\documentclass[preprint]{aastex}

\newcommand{\Msini}{\mbox{$M \sin i$}}

\newcommand{\Msun}{\mbox{$M_{\sun}$}}
\newcommand{\Mjup}{\mbox{$M_{Jup}$}}

\newcommand{\perone}{\mbox{$^{-1}$}}

\newcommand{\etal}{et al.}
\newcommand{\eg}{e.g.}
\newcommand{\ie}{i.e.}

\newcommand{\msyr}{\hbox{m~s$^{-1}$~yr$^{-1}$}}
\newcommand{\kms}{\hbox{km~s$^{-1}$}}

\newcommand{\htwoo}{{\hbox{H$_2$O}}}   
\newcommand{\meth}{{\hbox{CH$_4$}}}   
\newcommand{\Kp}{\mbox{$K^{\prime}$}}
\newcommand{\Ks}{\mbox{$K_S$}}
\newcommand{\degs}{\mbox{$^{\circ}$}}

\slugcomment{ApJ, in press}
\shorttitle{Liu \etal}
\shortauthors{Very Close L-Dwarf Companion to HR~7672}

\begin{document}

\title{Crossing the Brown Dwarf Desert Using Adaptive Optics:\\A Very
Close L-Dwarf Companion to the Nearby Solar Analog
HR~7672}

\author{\sc Michael C. Liu\altaffilmark{1}} 
\affil{Institute for Astronomy, University of Hawai`i, 2680 Woodlawn
Drive, Honolulu, HI 96822}
\altaffiltext{1}{Beatrice Watson Parrent Fellow.}
\email{mliu@ifa.hawaii.edu}


\author{\sc Debra A. Fischer, James R. Graham, James P. Lloyd, Geoff W. Marcy}
\affil{Department of Astronomy, 601 Campbell Hall, University of
California, Berkeley, CA 94720}


\author{\sc R. Paul Butler}
\affil{Department of Terrestrial Magnetism, Carnegie Institution of
Washington, 5241 Broad Branch Road, NW, Washington DC, 20015-1305}

\begin{abstract}

\noindent We have found a very faint companion to the active solar
analog HR~7672 (HD~190406; GJ~779; 15 Sge).  Three epochs of high
resolution imaging using adaptive optics (AO) at the Gemini-North and
Keck~II Telescopes demonstrate that HR~7672B is a common proper motion
companion, with a separation of 0\farcs79 (14~AU) and a 2.16~\micron\
flux ratio of 8.6~mags. Using follow-up $K$-band spectroscopy from Keck
AO+NIRSPEC, we measure a spectral type of L4.5$\pm$1.5.  This is the
closest ultracool companion around a main sequence star found to date by
direct imaging.  We estimate the primary has an age of
1--3~Gyr. Assuming coevality, the companion is most likely substellar,
with a mass of 55--78~\Mjup\ based on theoretical models.  The primary
star shows a long-term radial velocity trend, and we combine the radial
velocity data and AO imaging to set a firm (model-independent) lower
limit of 48~\Mjup.  In contrast to the paucity of brown dwarf companions
at $\lesssim$4~AU around FGK dwarfs, HR~7672B implies that brown dwarf
companions do exist at separations comparable to those of the giant
planets in our own solar system.  Its presence is at variance with
scenarios where brown dwarfs form as ejected stellar embryos.  Moreover,
since HR~7672B is likely too massive to have formed in a circumstellar
disk as planets are believed to, its discovery suggests that a diversity
of physical processes act to populate the outer regions of exoplanetary
systems.
\end{abstract}


\section{Introduction}

Radial velocity surveys find that about 6\% of solar-type FGK stars
harbor planets within 4~AU, with $M \sin i$ spanning 0.25--15~\Mjup. In
contrast, the same surveys find the incidence of more massive substellar
companions (15--80~\Mjup) at these radii is $\lesssim$1\%, even though
such objects are much easier to detect.  The evidence for this ``brown
dwarf desert'' first emerged from radial-velocity surveys in the late
1980's and early 90's \citep{1988ApJ...331..902C, 1989ApJ...344..441M,
1995Icar..116..359W}. Their modest precisions ($\sim$300~m/s) were
sufficient to detect brown dwarfs inside 3~AU, but only a single
candidate was found.\footnote{This object was the companion to HD~114762
with $\Msini = 11~\Mjup$ \citep{1989Natur.339...38L}. The primary star
is likely seen pole-on so its companion was believed to be stellar
\citep{1991ApJ...380L..35C}. Recent AO imaging has resolved a third
component, a late-type dwarf at 91~AU \citep{lloyd00}.  This raises the
possibility of dynamical interactions causing the stellar rotation axis
to be mis-aligned with the companion orbital axis \citep{pat01}.  Hence,
HD~114762B may in fact be a substellar companion.} Current
high-precision ($\sim$10~m/s) radial velocity programs have identified
about a dozen close companions with $M \sin i = 15-60~\Mjup$
\citep{1997abos.conf..313M}. However, the majority of these have been
found by {\sl Hipparcos} to be unresolved astrometric binaries with
M-dwarf companions, and the remainder are unlikely to be brown dwarfs
\citep{2000A&A...355..581H, zuck01, 2001A&A...372..935P}.  This paucity
of objects stands in stark contrast to the abundance of free-floating
brown dwarfs in the field \citep[e.g.][]{1999ApJ...521..613R,
2000AJ....120..447K} and in young clusters down to very low masses
\citep[e.g.][]{1998A&A...336..490B, 2000ApJ...540.1016L,
2000ApJ...541..977N, liu01a}.

An unanswered observational question is whether the brown dwarf desert
exists at $\gtrsim$4~AU, outside the region of current radial velocity
surveys.  A few L and T-dwarf companions have been found around FGK
dwarfs from the 2MASS imaging survey at very wide separations
(250--2500~AU) \citep{2000ApJ...531L..57B, 2001AJ....121.3235K,
wils01}. The statistics are currently small, but this suggests the brown
dwarf desert does not exists at very large separations
\citep{2001ApJ...551L.163G}.  These companions might have formed during
the fragmentation of the same natal molecular core as the primary. Given
their mass and large separation, it is unlikely that they formed in a
circumstellar disk, as planets are believed to form.

Little is known about the frequency of substellar companions at
$\approx$4--50~AU.  In our solar system, this is the domain of giant
planets and the Kuiper Belt. Hence, probing these separations around
other stars can test our understanding of formation processes in the
outer regions of planetary systems. There are clues that massive planets
and/or brown dwarfs do exist at these radii: half of stars with planets
have systematic long-term trends in their radial velocities due to
unseen companions \citep{2001ApJ...551.1107F}.  The orbital periods are
much longer than the observing baselines, and hence the masses are
poorly constrained.  Because the periods are many years and/or decades,
relying on radial velocities alone will require a long time to determine
the companion masses.  However, adaptive optics (AO) imaging with large
ground-based telescopes can probe the physical nature of these
companions and lend insight into the mass distribution of substellar and
planetary companions.

Thus, ground-based AO imaging provides a key capability for finding
substellar companions to main sequence stars, providing sensitivity at
radii outside radial velocity searches but closer than ordinary imaging
surveys.  Recently, \citet{els01} have found a common proper motion
companion to the extrasolar planet star GL~86.  Their coronagraphic AO
imaging finds a companion at a separation of 1\farcs7 (19~AU). The
companion is among the coolest known, probably at the transition between
L and T-type objects based on IR photometry.

Aside from implications for understanding the planetary formation
process, brown dwarf companions to main sequence stars are interesting
in their own right. Since a brown dwarf does not have a stable source of
internal energy, its age and mass are degenerate for a given spectral
type --- young lower mass BDs can have same temperature as older higher
mass BDs.  By finding brown dwarf companions to main sequence stars, we
can break this degeneracy by measuring the age of the primary and
assuming the components are coeval.  Combined with theoretical models
for the cooling history of substellar objects, we can determine masses
for brown dwarfs.

Here we report the discovery and characterization of a common proper
motion companion to HR~7672 (HD~190406, 15~Sge, GJ~779) using the AO
systems at the Gemini-North and Keck~II Telescopes.  The star lies at a
distance of 17.7~pc and has a spectral type of
G1V. \citet{1996A&ARv...7..243C} classified it as a solar analog, and
its higher level of activity than the Sun suggests a young age. Our
observations and astrometry of HR~7672B appear in \S~2. We present
temperature and mass determinations for the companion based on near-IR
AO photometry, spectroscopy, and radial velocity data in \S~3.  In \S~4,
we summarize our results and offer some implications for the formation
of substellar objects around solar-type stars.


\section{Observations and Astrometry}

\subsection{Imaging}

We first imaged HR~7672 on 24 June 2001 UT using the Gemini-North 8.1-m
telescope with the Hokupa'a AO system \citep{hokupaa96}.  The associated
imaging camera QUIRC has a scale of $19.98\pm0.08$~mas~pixel\perone.
Conditions were non-photometric with variable transmission due to patchy
cloud cover.  We obtained data in the \Kp-band filter (1.9--2.3~\micron;
\citealp{1992AJ....103..332W}), taking both short and long exposure
images to cover a large dynamic range.  In the latter, the primary is
saturated.  The AO images have a 0\farcs10 FWHM core, with much of the
light in a broader seeing-limited halo, typical for images with only
modest AO correction.  We identified a faint point source close to the
primary.\footnote{We also detected three faint sources farther away.  We
had previously observed this field using the AO system on the Shane~3-m
telescope at Lick Observatory \citep{1999SPIE.3762..194B}.  $K$-band
images taken with the science camera IRCAL \citep{2000SPIE.4008..814L}
on 02~August~1999~UT showed sources at ($d$, PA) = (10\farcs0,
142\degs), (7\farcs9, 178\degs), and (5\farcs6, 256\degs), all around
8--10~mags fainter than the primary.  All three are detected in the
Gemini AO image and are found to be background objects.}

To check for common proper motion, we obtained second and third epoch
imaging using the AO system \citep{2000PASP..112..315W} on the Keck~II
10-m Telescope on 22~Aug and 10~Dec 2001~UT.  We used the slit-viewing
camera of the facility near-IR spectrograph NIRSPEC
\citep{1998SPIE.3354..566M} and a narrow-band Br$\gamma$ filter
(2.155--2.175~\micron).  The camera has a pixel scale of
$16.74\pm0.05$~mas~pixel\perone.  Conditions were photometric,
and the AO-corrected images have a 0\farcs05 FWHM.  The companion was
easily seen in individual images.

\begin{figure}[t]
\vskip -1in
\centerline{\includegraphics[width=4in,angle=90]{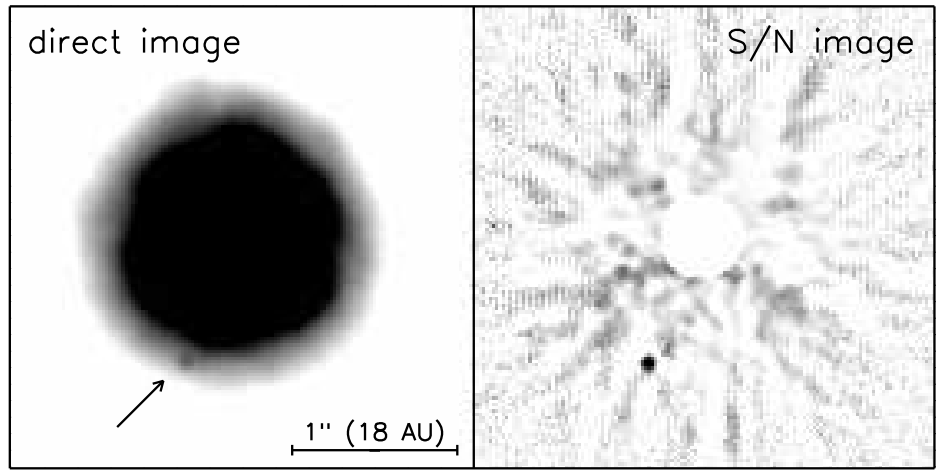}}
\vskip -6ex
\caption{{\em Left:} \Kp-band discovery image of HR~7672B from
Gemini-North AO. The companion is seen as a faint object in the halo of
the primary. The orientation is as on the sky, North up and East
left. {\em Right:} S/N image of the same data, constructed by unsharp
masking the original data and then dividing by a radial profile whose
amplitude follows the radial noise profile. The result is an image with
uniform noise over the entire field --- the companion is
well-detected. For both panels, the color stretch is
linear.\label{fig-gemini}}
\end{figure}

The imaging data were reduced in a standard fashion.  We subtracted an
average bias from the images.  We constructed flat fields either from
images of the lamp-illuminated interior of the dome (for the Gemini
data) or from images of twilight sky (for the Keck data).  Then we
created a master sky frame from the median average of the
bias-subtracted, flat-fielded images and subtracted it from the
individual images.  Images were registered and stacked to form a final
mosaic, though the astrometric measurements are actually done on the
individual images.  Figure~\ref{fig-gemini} shows the Gemini AO
discovery image.

To determine the separation and position angle for the companion, we
measured the individual reduced images and averaged the results, with
the uncertainties determined from the standard errors.  These errors
were then added in quadrature to the errors in the calibration of the
cameras' pixel scales and orientations. For the Keck AO data, the
astrometry was straightforward given the high quality of the data: the
companion was well-separated from the primary so we simply computed the
centroid of the two components.
We also used the Aug 2001 data to measure a 2.16~\micron\ flux ratio of
$8.62\pm0.07$~mags.

For the lower quality Gemini AO data, more care was required.  In the
short exposure images, where the primary was unsaturated, the companion
was detected only the final stacked mosaic and only at low S/N.  Hence
the resulting astrometry would have had a large uncertainty, and we
would not have been able to directly gauge the errors.  Instead, we
derived astrometry from the deep images, where the core of the primary
was saturated inside of 0\farcs1--0\farcs2 radius but the rest of the
primary was detected at very high S/N.  To measure the primary's
position, we cross-correlated the individual saturated images with an
unsaturated short exposure image. We then removed the primary by
subtracting its azimuthally averaged radial profile and measured the
companion's centroid.
We then averaged the results to get the separation and PA.  We verified
that these results are robust by (1) doing the measurements with several
different PSFs and (2) generating artificial data sets of saturated
images from the short exposure images. Both of these confirm our quoted
measurement accuracy.

Table~\ref{table-astrometry} presents our astrometry for the companion.
HR~7672 has a high proper motion, ($-$394.07, $-$406.42)~$\pm$~(0.63,
0.64) mas/yr, allowing us to confirm if the companion is physically
associated relatively quickly.  Nearly half a year passed between the
first and third epochs of imaging, and Table~\ref{table-astrometry}
shows that if the object were a background object, its separation and PA
would have changed significantly.  The observed separation and PA are
7$\sigma$ and 26$\sigma$ discrepant from being a background object,
respectively.  In fact the first two epochs of imaging, spaced only
2~months apart, would have sufficed.  Moreover, the second and third
epoch data by themselves indicate the companion is not a background
object; since these both came from the same instrument and filter, this
eliminates any concern about systematic error due to using different
instruments.
Hence we conclude HR~7672B is physically associated with HR~7672.

\subsection{Spectroscopy}

\begin{figure}[t]
\hskip 0.5in
\centerline{\includegraphics[width=3in,angle=90]{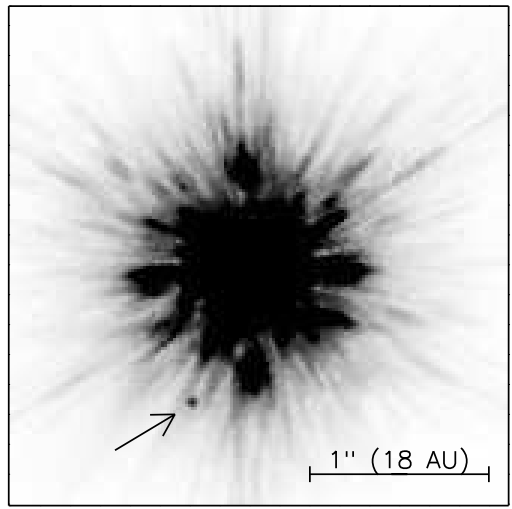}}
\vskip -3ex
\caption{$K$-band image of HR~7672B obtained with Keck AO. The color
scale is linear, and the image is the same size and orientation as in
Figure~\ref{fig-gemini}.
\label{fig-keck}}
\end{figure}

We obtained a spectrum at Keck using NIRSPEC with the AO system via
service observing.  We obtained two 300~s exposures on 31 Aug 2001 UT in
low resolution mode with the NIRSPEC-7 blocking filter and the
0\farcs072 slit.  The slit was oriented at the parallactic angle. (In
contrast to seeing-limited near-IR observations, atmospheric dispersion
with these nearly diffraction-limited images was not negligible.)  The
object was dithered on the slit between exposures.  Conditions appeared
to be photometric, and the AO-produced images had 0\farcs06 FWHM as
measured from the acquisition images taken with the slit-viewing camera.
A nearby A0V star was observed immediately afterward to calibrate the
telluric and instrumental absorption profile.  Images of an internal
flat-field and arc lamps were obtained when finished with the calibrator
star.  For comparison with the Gemini AO image, Figure~\ref{fig-keck}
shows the $K$-band image of the system obtained from the NIRSPEC
slit-viewing camera.

The spectra were reduced using custom IDL scripts.  In NIRSPEC, the raw
images on the detector are curved in both the spectral and spatial
direction.  After subtracting a dark frame and dividing by a flat field,
the individual images were rectified using traces of the arc lamp lines
and the object spectra.  Since the primary is very bright, the light
from the PSF halo and diffraction features fill much of the slit.  We
removed the telluric emission by fitting blank regions at the edges of
the slit and subtracting the result from the entire image.  To extract
the companion spectrum, we then used a 3-pixel wide region. To remove
the remaining primary light, at each wavelength we fit for the
contamination using pixels above and below the companion spectrum.  This
amounted to $\approx30$\% of the light in the extraction aperture.

\begin{figure}[t]
\vskip -0.5in
\hskip 0.5in
\includegraphics[width=4in,angle=90]{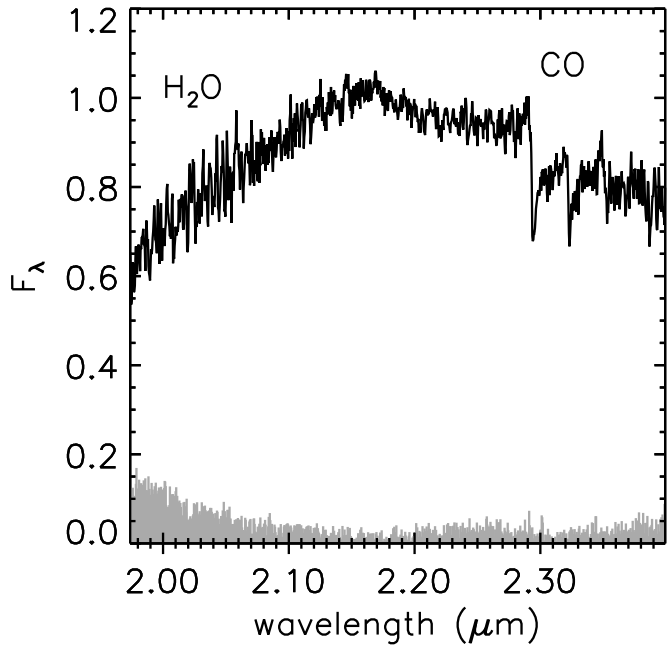}
\caption{$K$-band spectrum for HR~7672B obtained with Keck
AO+NIRSPEC. The spectrum has been scaled to an arbitrary constant. The
lines at the bottom are the 1$\sigma$ errors as determined from the
scatter in the individually extracted spectra.  The companion shows
strong \htwoo\ absorption on the blue side and strong CO absorption in
the red. \label{fig-spectra}}
\end{figure}

To calibrate for the relative throughput of the atmosphere and
instrument, we divided the extracted spectra by the spectra of the A0V
calibrator star and then multiplied by a 9720~K blackbody to restore the
true shape of the continuum.  For this, we interpolated over the
intrinsic Br$\gamma$ absorption in the calibrator star.  Wavelength
calibration was done with spectra of argon and neon lamps.  The spectral
resolution ($\lambda/\Delta{\lambda}$) of the final spectra was 1400, as
measured from the FWHM of the arc lines.  The final spectrum is plotted
in Figure~\ref{fig-spectra}, with the 1$\sigma$ uncertainties determined
from the standard error of the two individual spectra.

\section{Results}

\subsection{Temperature of HR~7672B}

In order to determine the spectral type of HR~7672B, we compare its
$K$-band spectrum with that of other very cool M and L-type objects.  We
use published spectra for (non-subdwarf) M~dwarfs from
\citet{2000ApJ...535..965L} and spectra from L~dwarfs from
\citet{geb01}, \citet{2001ApJ...548..908L}, and
\citet{2001AJ....121.1710R}; hereinafter, we refer to this compilation
as the ``calibration sample.'' For the L~dwarfs, we use the spectral
types assigned by \citet{geb01}, which are based on several optical and
near-IR spectral indices; for most objects, their spectral types agree
well with previous typing based on optical spectroscopy alone.
Qualitatively, the spectrum of HR~7672B shows strong blueward absorption
indicative of \htwoo\ and a strong CO 2.3~\micron\ break. These are
characteristic of very late-type M dwarfs and L dwarfs.  In addition, no
absorption is seen in 2.2~\micron\ from Na~I; this feature is seen in
late M~dwarfs \citep{1994MNRAS.267..413J, 2000ApJ...535..965L} but
disappears in L~dwarfs.  These properties suggest HR~7672B is an
L~dwarf.

Among L~dwarfs, there are noticeable variations in the published
$K$-band spectra for a given subtype.  The origins of these variations
are unknown, but could be due to heterogeneity in time variability,
photospheric dust properties, and/or surface gravities. Hence, a
qualitative (eyeball) comparison between HR~7672B and the published
spectra does not yield a clear and unique match.  To quantitatively
assign a spectral type, we compute several spectral indices: two indices
which measure \htwoo\ absorption on the blue side of the bandpass
(\htwoo-C from Burgasser \etal, and \htwoo(2.0) from Geballe \etal), an
index which tracks the slope of the 2.1--2.3~\micron\ continuum
(\meth(2.2) from Geballe \etal)\footnote{This index is primarily
designed to track the depth of 2.2~\micron\ absorption in T~dwarfs due
to \meth.  However, it is also a reasonable diagnostic for L~dwarfs, as
discussed by Geballe \etal}, and a CO bandhead index of our own
construction ($R_{CO}$ = the ratio of 2.30--2.32~\micron\ flux to
2.26--2.28~\micron\ flux).  We degrade our NIRSPEC spectra to a spectral
resolution comparable to the published spectra before measuring the
indices.  To estimate our measurement errors, especially given the
possibility that there is some residual contamination in our spectra
from the light of the primary, we measure the two extracted spectra
individually and plot the resulting range.

\begin{figure}[t]
\vbox{\hskip 0.3in
\includegraphics[width=2.8in,angle=0]{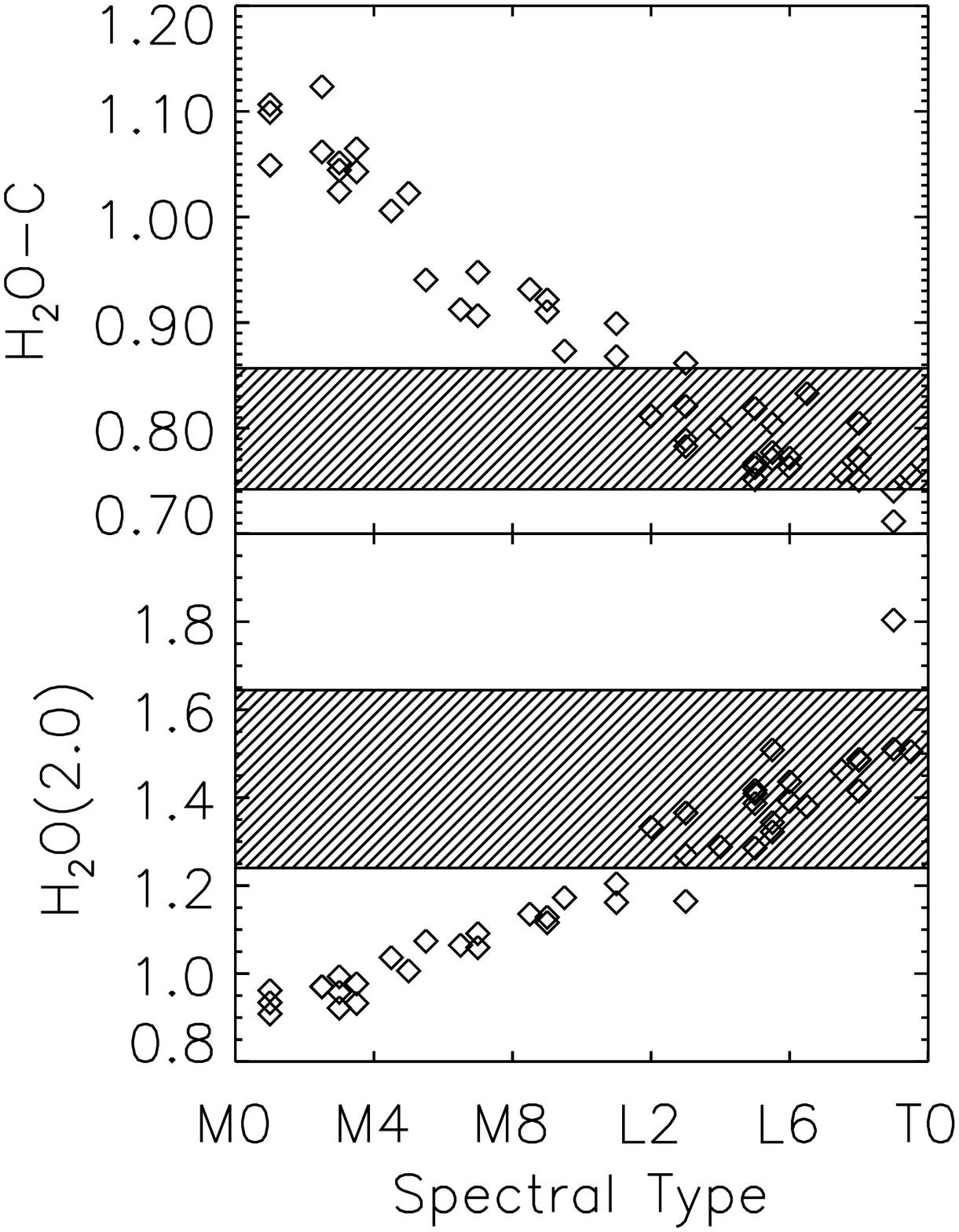}
\hskip 0.5in
\includegraphics[width=2.8in,angle=0]{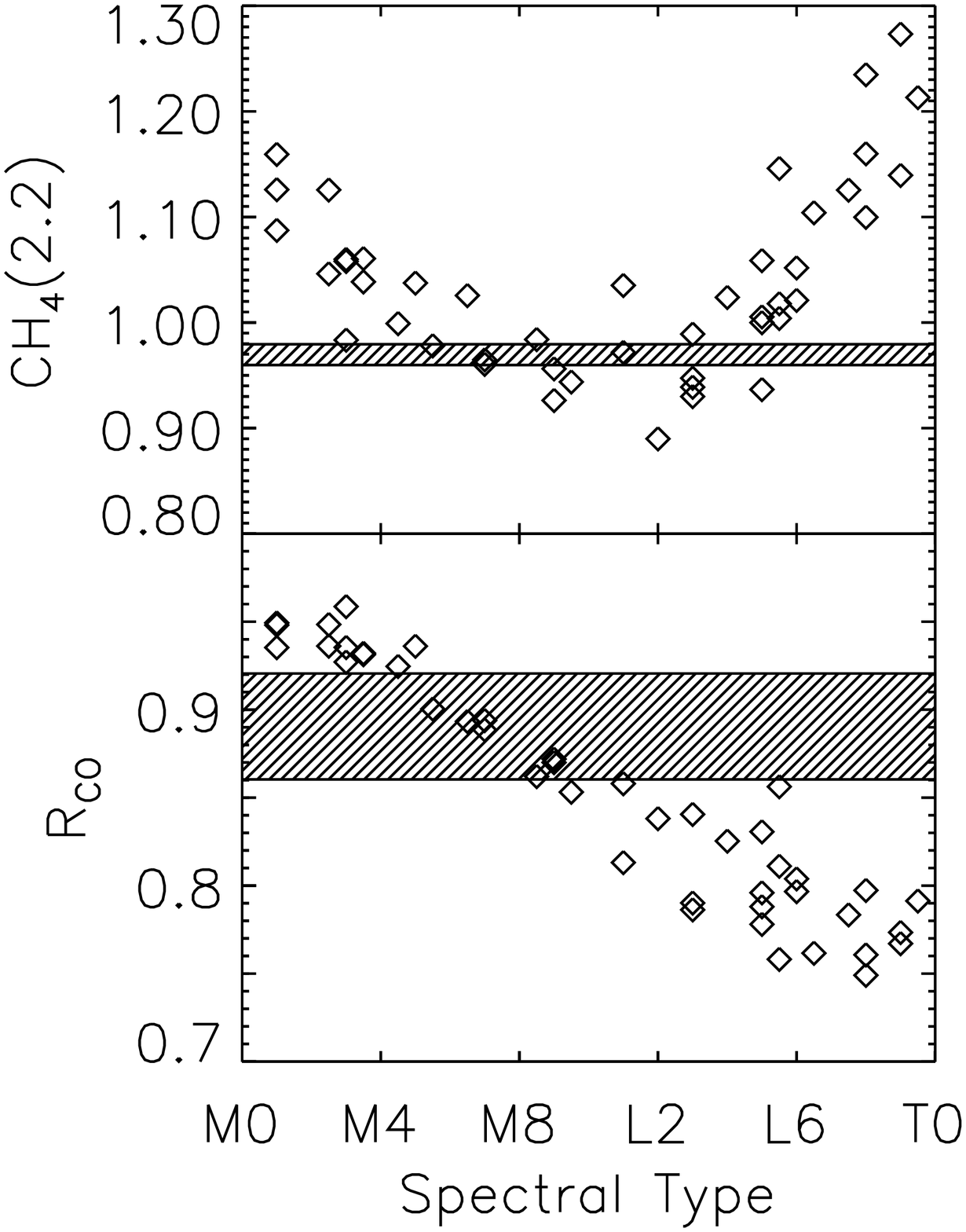}}
\vskip 4ex
\caption{Spectral classification of HR~7672B based on molecular spectral
indices for very cool atmospheres (see \S~3 for references).  The
diamonds are objects from the literature with known spectral types.  The
hatched bands represent the HR~7672B measurements. The indices on the
left trace the increasing \htwoo\ absorption at the blue edge of the
bandpass; these indicate HR~7672B has spectral type of L2 or later. The
ones on the right track the shape of the continuum and the depth of CO
absorption; these suggest a type no later than L6.  The CO absorption
appears to be anomalously weak. \label{fig-indices}}
\end{figure}

The results are shown in Figure~\ref{fig-indices}.  The strength of the
\htwoo\ absorption suggests a spectral type later than L2, while the
continuum \meth\ indices point to a type no cooler than L6.  The CO
index is discrepant; the strength of the absorption is weaker than most
L dwarfs, and is more typical of mid/late M dwarfs.  To disentangle this
uncertainty, we also use the absolute $K$-band magnitude of HR~7672B to
estimate the spectral type.  \citet{2000AJ....120..447K} have compiled a
sample of 24 nearby late~M and L~dwarfs with measured parallaxes to
generate a calibration for spectral type with \Ks-band (2.0--2.3~\micron)
absolute magnitude:
\begin{equation}
M_{\Ks} = 10.450 + 0.127(subclass) + 0.023(subclass)^2
\end{equation}
where $subclass=-1$ for M9V, 0 for L0V, 1 for L1V, etc.  The scatter
about the fit is $\approx$1~subtype.  HR~7672 has a $V$-band magnitude
of 5.80~mag, a distance of 17.7~pc, and a spectral type of G1V. Using
$V-K$ colors for dwarfs from \citet{1988PASP..100.1134B}, this means the
primary has $M_K=3.1$~mag.  For G~stars, we use synthetic photometry of
spectra from \citet{1998PASP..110..863P} to determine that there is a
negligible difference between the $K$ and \Ks-band magnitudes.  For
HR~7672B, we measure a 2.16~\micron\ flux ratio of 8.6~mags. We use
synthetic photometry of the calibration sample to find that
\hbox{(\Ks--[Br$\gamma$])}~ $\approx0.05-0.10$~mags for L~dwarfs, so the
companion has $M_{\Ks}=11.8\pm0.1$~mags, which gives a spectral type of
L5.5.  Given the relatively small L~dwarf sample and the scatter about
the fit, spectral types between L4 and L6 could be accommodated.

We adopt a final spectral type of L4.5$\pm$1.5.  This is based on the
intersection of the results from the spectral indices, which suggest L2
to L6, and the $K$-band absolute magnitude, which points to L4 to L6.
The spectral typing (and hence the mass estimate) could be improved in
the future by obtaining spectra in the $H$-band and $J$-band; the latter
would require good seeing conditions given the currently available AO
systems on $>$8-m telescopes.  To convert to effective temperature, we
use the scale produced by \citet{burg01}, which is intermediate between
those proposed by \citet{1999ApJ...519..802K,2000AJ....120..447K} and
\citet{2000ApJ...538..363B}.  This choice gives $T_{eff}=1510-1850$~K
for HR~7672B.

\subsection{Mass of HR~7672B}

\subsubsection{From Spectroscopy}

Not all L~dwarfs are brown dwarfs (i.e. substellar).  Current evidence
suggests that the substellar boundary occurs around spectral types of
L2--L4 \citep{1999ApJ...519..802K, 2000ARA&A..38..485B}.  With a
spectral type of L4.5$\pm$1.5, HR~7672B lies at or below this boundary.
A useful point of comparison is GD~165B, which has a spectral type L4
and an age estimate of 1.2--5.5~Gyr from its primary star. A detailed
analysis by \citet{1999ApJ...519..834K} finds that GD~165B is probably a
brown dwarf.  Detecting lithium absorption in HR~7672B would be an
unambiguous sign that the companion is substellar.  However, given its
small separation and large brightness difference
($\Delta{I}\approx13.5$~mag), this is unlikely to be feasible any time
soon.

\subsubsection{From Radial Velocities}

HR~7672 was included in the original Lick radial velocity survey and was
found to have a long-term radial velocity trend (acceleration).  We can
use this data to determine the minimum possible companion mass strictly
from dynamics.  The primary's radial acceleration is related to the
companion mass by
\begin{equation}
M_{comp} = 5.34\times10^{-6}\ \Msun 
  \left(\frac{d}{\rm pc}\ \frac{\rho}{\rm arcsec}\right)^2 
  \left|\frac{\dot{v_r}}{\msyr}\right|\ 
  F(i,e,\omega,\phi)
\end{equation}
where $d$ is the distance to the primary from Earth, $\rho$ is the
observed angular separation of the companion, and $\dot{v_r}$ is the
radial acceleration \citep{1999PASP..111..169T}.  $F(i,e,\omega,\phi)$
is a function which depends on the orbital parameters (inclination $i$,
eccentricity $e$, longitude of periastron $\omega$, and orbital phase
$\phi$) and has a minimum value of $\sqrt{27}/2$.  From the original
Lick radial velocity survey, \citet{1999ApJ...526..890C} reported a
long-term radial velocity trend (acceleration) of $-24\pm0.6$~\msyr\
over 11.4~yrs.  This gives a minimum companion mass of $66\pm2$~\Mjup\
(but see below).

As discussed by \citet{1999PASP..111..169T}, it is possible to evaluate
the companion mass in a statistical fashion given the observed
acceleration and angular separation at a single epoch.  Basically, the
approach is to generate a Monte Carlo realization of
$F(i,e,\omega,\phi)$ by choosing randomly distributed values for the
orbital parameters.  The resulting probability distribution function
(PDF) can be used to set statistical upper limits on the companion mass.
(Note that the PDF has a long tail at high masses which arises from
highly improbable orbits. We adopt an upper mass cutoff of ~1~\Msun\
based on physical plausibility and then re-normalize the PDF.)  Applying
this technique to HR~7672B, we find a maximum companion mass of
0.11~\Msun\ at the 68\% confidence limit (1$\sigma$), a relatively weak
constraint.  Note that the most probable companion mass from this
approach is the same as the minimum possible mass based on the observed
$\dot{v_r}$ and $\rho$, namely 66~\Mjup.

The approach of Torres (Eqn.~2) makes use of the instantaneous radial
velocity slope, $\dot{v_r}$.  However we have radial velocity
measurements from the last 13 years.  This duration places a better
constraint on the orbit, given the difficulty of accurately measuring
the instantaneous slope.  In particular, the velocity slope at the
current epoch may be slightly different than the average slope during
the entire monitoring, which is what is listed in
\citet{1999ApJ...526..890C}.

\begin{figure}[t]
\hskip -0.25in
\includegraphics[width=2.5in,angle=90]{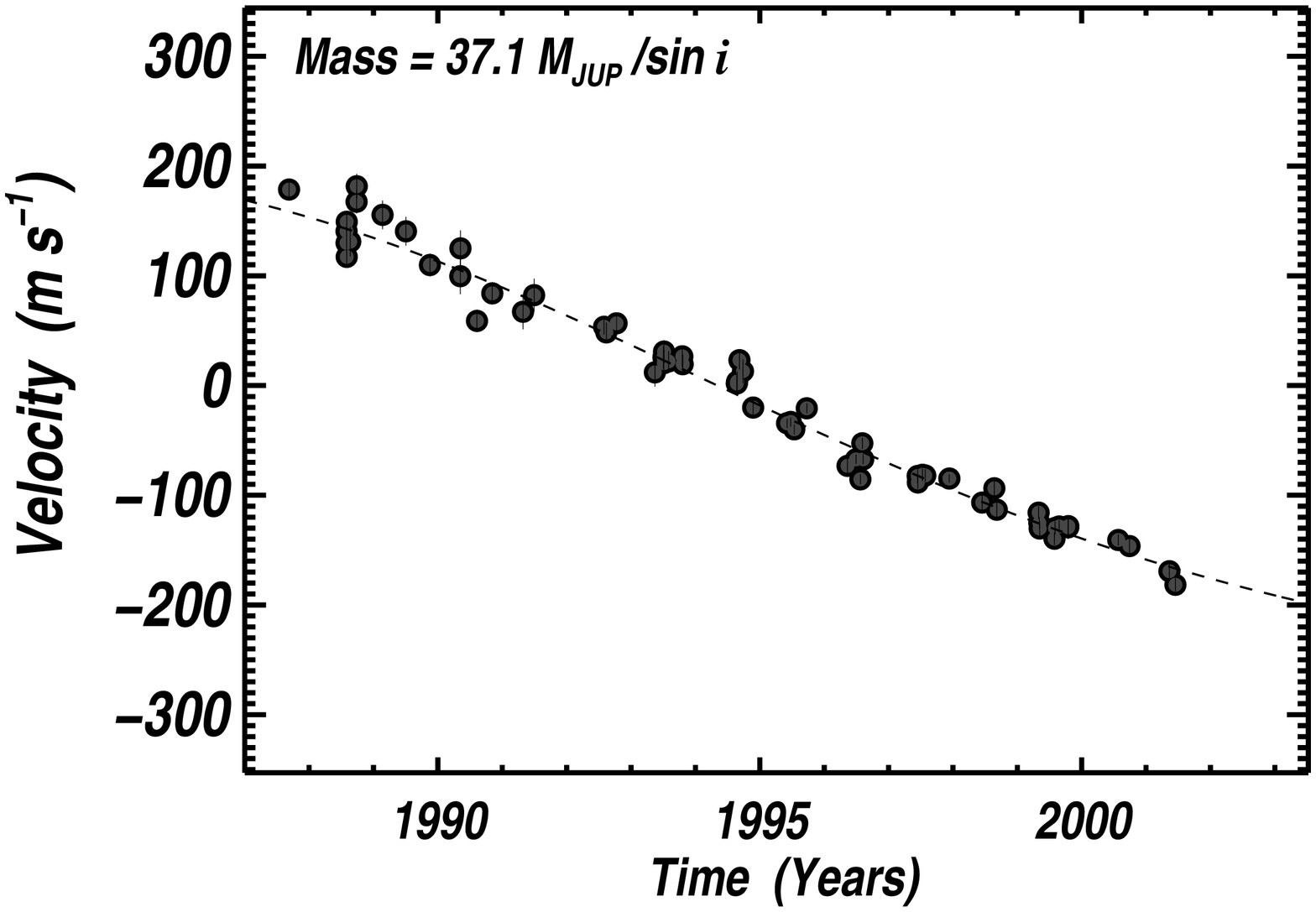}
\includegraphics[width=2.5in,angle=90]{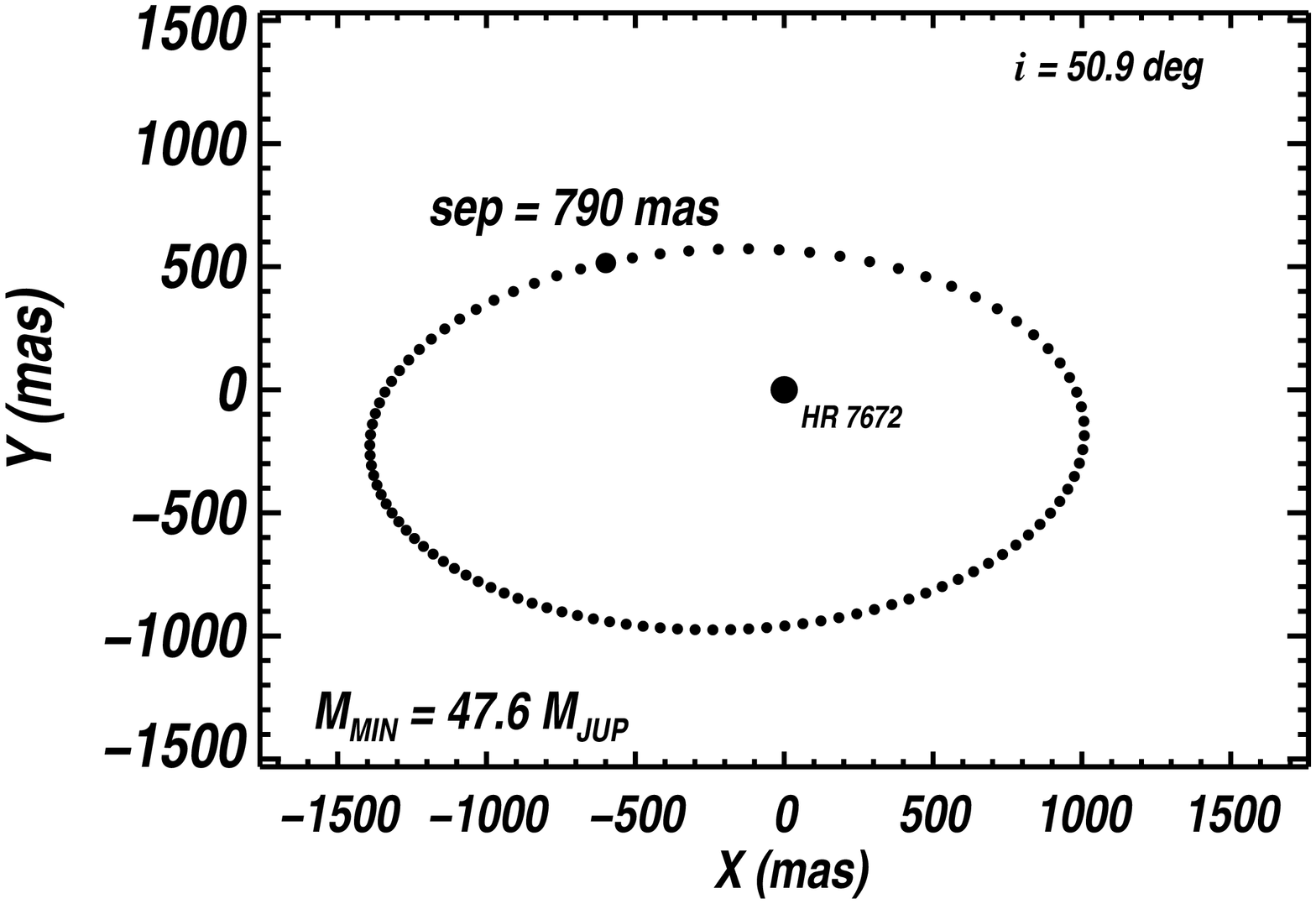}
\caption{One plausible orbit for HR~7672B, of the many possible ones,
which is consistent with the radial velocity and AO data.  This orbit
implies a minimum mass for the companion of 48~\Mjup\ and has a period
of 100~yr, eccentricity of 0.3, and inclination of 51\degs. {\em Left:}
Measured velocities (dots) and the velocity curve of the plausible orbit
(dashes).  {\em Right:} The same orbit but showing the motion of the
companion in the plane of the sky relative to the primary, yielding a
separation of 0\farcs79 at the current epoch, as observed.  Orientation
on the sky is arbitrary.
\label{fig-orbit}}
\end{figure}

To find the most accurate minimum companion mass, we constructed a
family of orbits that fit both the radial velocity data and the observed
separation of 0\farcs79.  Note that with only two observables, namely
the separation and velocity slope, no unique orbit can be established.
Nonetheless, we searched the orbital parameters for the lowest possible
companion mass consistent with the data.  For these minimum-mass orbits,
the instantaneous slope in the radial velocities was typically 25\% less
than the average during the past 13 years.  Such orbits imply a 25\%
lower mass for the companion, as given by Eqn.~2. This difference can be
seen in the plotted velocities (Figure~\ref{fig-orbit}), as the most
recent data indicate a slight flattening in the velocities.

One such minimum-mass orbit is shown in Figure~\ref{fig-orbit}.  The
companion has mass of 48~\Mjup\ with a 100~yr orbital period, a
semi-major axis of 21~AU, an eccentricity of 0.3, and an inclination of
51\degs.  This hypothetical orbit fits the measured radial velocities of
the primary within errors, and it would place the companion at a
separation of 0\farcs79 at the current epoch, as observed.  No orbits
are consistent with a companion mass less than 48~\Mjup.
Therefore we conclude that HR~7672B has a mass
of at least 48~\Mjup.

\subsubsection{From Age of the Primary}

We can also estimate the companion mass by determining the age of the
primary.  Assuming the system is coeval, we can then use the measured
temperature of HR~7672B with theoretical models to estimate its
mass. Main sequence G~stars can be age dated using a variety of
indicators, many of which are based on the anti-correlation of stellar
activity with age. None of these indicators are perfectly correlated
with the age of the star (for reasons which remain to be understood),
but all follow general trends.

\begin{enumerate}

\item Absolute $V$-band magnitude: with $M_V=4.56$, the primary is on the
main sequence, which sets an age greater than the zero-age main
sequence, $\approx$100~Myr.

\item Rotation period: The rotation of solar-type stars is believed to
increase as they age in a power-law fashion, $P_{rot} \sim
t^{\alpha}$. The value for $\alpha$ has been suggested to be 0.5
\citep{1972ApJ...171..565S} or $1/e$ \citep{wal91}.
\citet{2001A&A...366..215M} measure a photometric period of 13.95~d for
HR~7672, which we adopt as the stellar rotation period.  Taking the Sun
as a reference point ($P_{rot} = 26$~days and $t=4.6$~Gyr) gives
an age of 0.9--1.4 Gyr.  \citet{1999A&A...348..897L} also have provided
a calibration of this relation using a sample from the {\sl Hipparcos}
catalog:
\begin{equation}
\log(t) = 2.667 \log(P) - 0.944 (B-V) - 0.309 [{\rm Fe/H}] + 6.530 .
\end{equation}
where $t$ is the age in Gyr and $P$ is the period in days.  For HR~7672,
$B-V=0.60$ \citep{1995ApJ...438..269B} and [Fe/H]=--0.07
\citep{1999A&AS..139...29G} so we find an age of 1.1~Gyr.

\item X-ray emission: Related to the decline in stellar rotation, the
X-ray emission of solar analogs declines with
age. \citet{1995A&A...294..515H} measured $\log(L_X/L_{bol})=-5.54$ from
ROSAT data.  \citet{1998PASP..110.1259G} provides an age calibration
based on scaling relations for stellar activity:
\begin{equation}
\log(L_X/L_{bol}) = -6.38 - 2.6\alpha\log(t/4.6) + \log[1 + 0.4(1-t/4.6)]
\end{equation}
where $t$ is the age in Gyr and $\alpha$ is the same as given above. We
have adopted the zeropoint of $-$6.38 from
\citet{1987ApJ...315..687M}. This gives an age of 0.8--1.3~Gyr.

\item Ca H+K emission: Chromospheric activity also decreases with
age. From Keck spectroscopy in 1997, we measured
$\log(R\arcmin_{HK})=-4.79$ for the primary.  As cited by
\citet{2001AJ....121.3235K}, \citet{don93, 1998csss...10.1235D} provide
an age calibration for this index
\begin{equation}
\log(t) = 10.725 - 1.334 R_5 + 0.4085 R_5^2  - 0.0522 R_5^3
\end{equation}
where $R_5 = 10^5 R\arcmin_{HK}$.  This gives an age of
2.6~Gyr.

\item Kinematics: Older stars tend to have higher space velocities due
to an accumulated history of dynamical interactions.  The
\citet{1991NSC3..C......0G} catalog lists $U=+40, V=-19, W=+9$~\kms\ for
HR~7672.  This lies just outside the young disk population defined by
\citet{1992ApJS...82..351L}.\footnote{Note that the Gliese \& Jahreiss
catalog and Leggett use the opposite sign conventions for $U$.}
\citet{1989PASP..101..366E} has approximated the division between old
and young disk at $\sim$1.5~Gyr.

\item Lithium absorption: Li is steadily burned in main sequence stars
with convective envelopes.  \citet{1985PASP...97...54S} measured a
lithium abundance of $N$(Li)=2.15.  This is weaker than Hyades stars of
the same spectral type \citep{1993ApJ...415..150T, 1993AJ....106.1059S},
which sets a lower limit of $\sim$600 Myr on the age of the primary.
Also, the abundance is more than early-type G stars in the $\sim$4.5~Gyr 
open cluster M67 \citep{1999AJ....117..330J}.

\end{enumerate}

To summarize, HR~7672 exhibits a higher level of chromospheric activity
than the Sun, pointing to a younger age.  The primary's kinematics
suggest it is an old disk star, while the lithium abundance points to an
age greater than Hyades stars. The star's rotation period, Ca~H+K
emission, and X-ray emission suggest ages of 1--3 Gyr, which we adopt as
a conservative age range for the primary.

\begin{figure}[t]
\vskip -0.25in
\hskip -0.5in
\includegraphics[width=2.6in,angle=90]{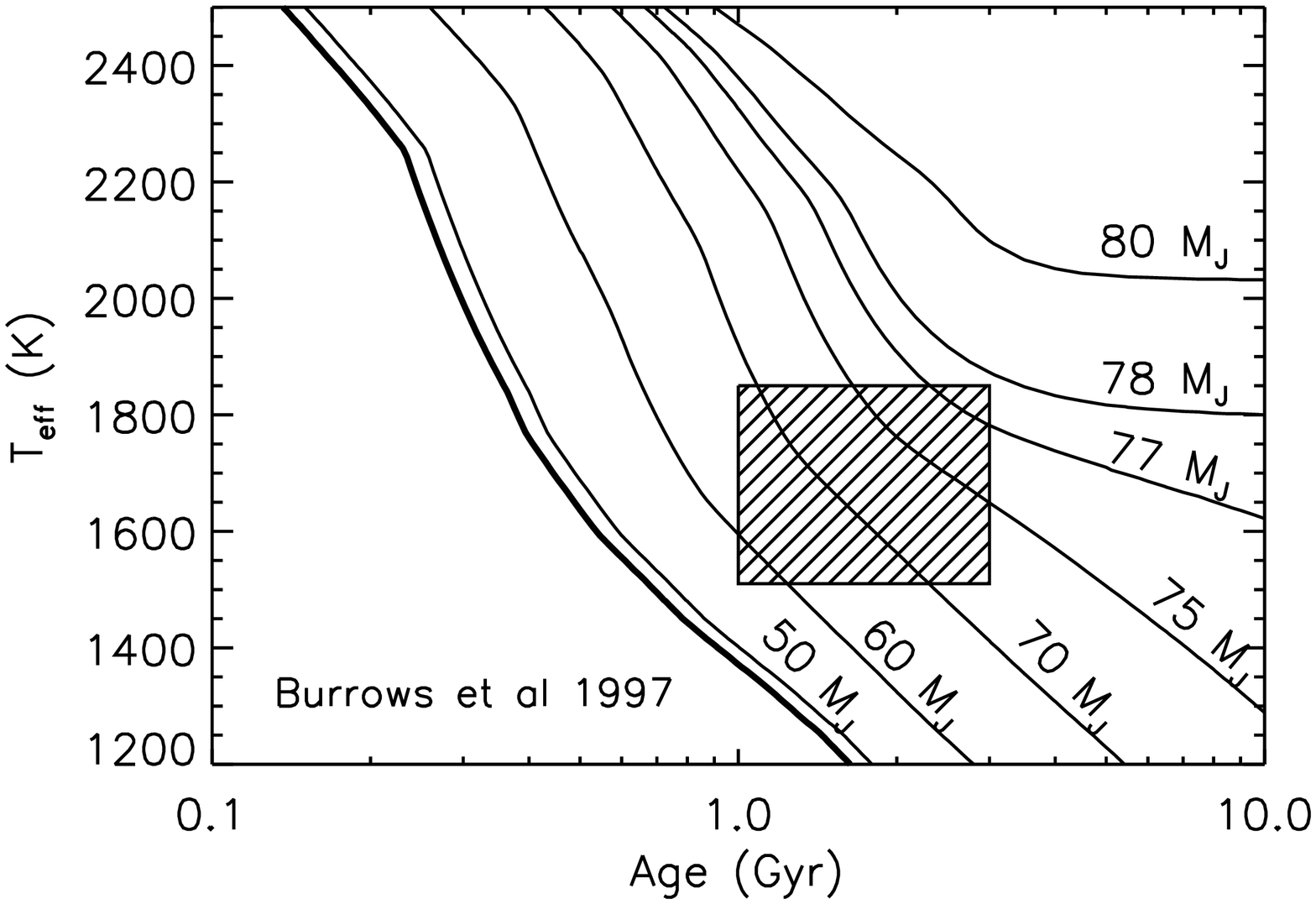}
\includegraphics[width=2.6in,angle=90]{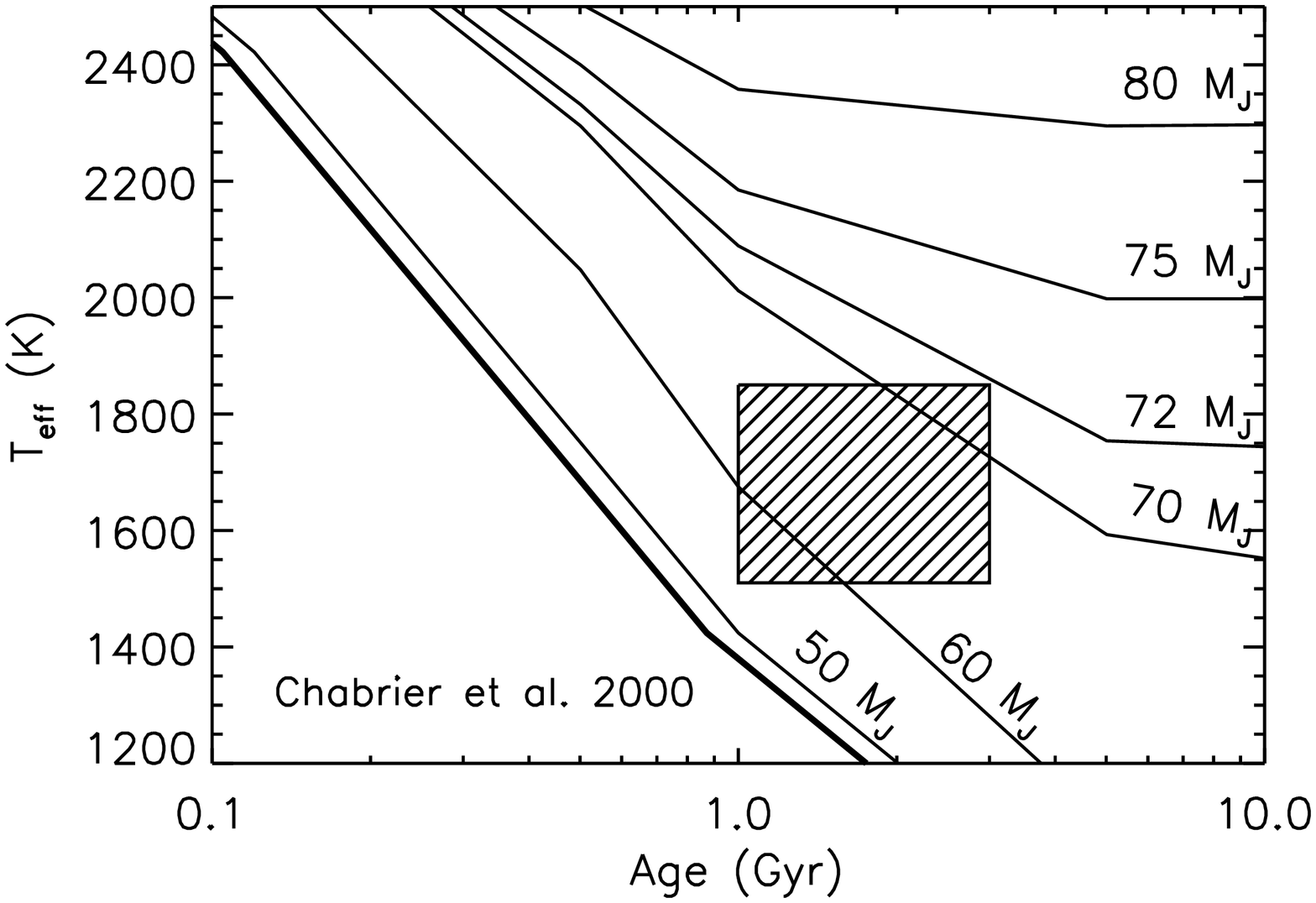}
\vskip -4ex
\caption{Mass determination for HR~7672B based on theoretical models
from \citet{1997ApJ...491..856B} and \citet{2000ApJ...542..464C}.  The
hatched region indicates the observational constraints, a combination of
the age of the primary and the spectral type of the companion.  The
inferred mass ranges from 55 to 78~\Mjup. The heavy line represents the
model-independent lower limit of 48~\Mjup\ from the radial velocity and
AO data.
\label{fig-models}}
\end{figure}

Assuming the system is coeval, we estimate the mass of HR~7672B using
theoretical models (Figure~\ref{fig-models}).  With an implied range of
55--78~\Mjup, the companion most likely lies below the minimum H-burning
mass.  (Note that we can also reverse the comparison and use the radial
velocity limit of 48~\Mjup\ with the models to find a lower limit of
$\sim$0.3--0.7~Gyr on the age of the primary, consistent with the
observational properties described above.)  This mass estimate could be
improved in the future by obtaining multi-band near-IR spectra to
improve the spectral typing.

\section{Conclusions}

We have found a common proper motion companion to the nearby solar
analog HR~7672 using AO imaging from the Gemini and Keck Telescopes. IR
spectra from Keck AO+NIRSPEC find the companion has a very cool
atmosphere, and its faint $K$-band magnitude also points to a very
late-type object: we estimate a spectral type of L4.5$\pm$1.5.  This
alone suggests the companion is sustellar, or right at the substellar
boundary.  A variety of data indicate the primary star is younger than
the Sun but older than the Hyades, with a likely age around 1--3~Gyr.
Using theoretical models for cooling of very low mass objects, the
inferred companion mass ranges from 55--78~\Mjup, under the assumption
that the two components are coeval.  The primary has a long-term radial
velocity acceleration, and we combine the radial velocity and AO data to
set a lower limit of 48~\Mjup\ on the companion mass strictly from
dynamical considerations.  At a separation of only 0\farcs79 (14~AU),
this is the closest ultracool (L or T-dwarf) companion to a main
sequence star found to date by direct imaging, in both angular and
physical separation (see compilation in \citealp{2001AJ....121..489R}).

For a random distribution of orbital eccentricities, $\approx$85\% of
orbits have a true semi-major axis which is 0.5--2~times that of the
observed separation.  Hence the semi-major axis of HR~7672B is likely to
be 7--28~AU, meaning an orbital period of 20--150~yrs. Orbital motion
could be detectable in a few years. This raises the possibility of
determining its orbital eccentricity, perhaps a important clue into its
formation history.  Furthermore, the combination of AO imaging with
continued radial velocity monitoring can be used to better constrain the
companion mass in advance of observing a full orbital period.

The formation and presence of close brown dwarf companions might inhibit
the formation of circumstellar disks, and hence planets.  Using radial
velocity data, \citet{1999ApJ...526..890C} set upper limits on any
planetary companion of 0.1--10~\Mjup\ at separations of 0.1--4~AU,
respectively.  Hence, we know the HR~7672 system does not have any
Jovian mass planets in its inner regions.
However, we know from the case of GL~86 that at least one brown dwarf
co-exists with an extrasolar planet \citep{els01}.  The existence of
HR~7672B and GL~86B indicates that brown dwarfs can form at separations
comparable to circumstellar disks.  (These objects are most likely
substellar, or right at the substellar boundary.  In the discussion
which follows, we assume for the sake of arugment that these are brown
dwarfs.)  A third possible example is the brown dwarf companion orbiting
HD~168443 \citep{2001ApJ...555..418M}, at a semi-major axis of 3~AU and
with a mass between 17.2~\Mjup\ (from the radial velocities) and
42~\Mjup\ (from limits on any astrometric wobble).  This system also
illustrates the gradual convergence of radial velocities and AO imaging,
closing the mass gap for discovery of planets and brown dwarfs by
indirect and direct means.

Formation scenarios for HR~7672B, and their connection with ones for
extrasolar planets, remain an open question.  Current semantic
convention designates planets as object below $\sim$15~\Mjup, with this
border motivated either by the deuterium-burning limit for substellar
objects or simply by the observed steep decline in frequency of such
objects in radial velocity surveys \citep{2001ApJ...555..418M}; more
massive objects are considered brown dwarfs.  HR~7672B would certainly
be considered in the brown dwarf regime based on its estimated mass.
\citet{2001AJ....122..432R} have suggested brown dwarfs form as stellar
embryos in small newborn multiple systems: they grow in mass by
accretion of infalling gas, but their growth is prematurely truncated by
ejection due to dynamical interactions.  The discovery of HR~7672B is at
variance with this scenario, since the ejection model discounts the
existence of close separation brown dwarfs around solar-type stars.

An additional criterion to distinguish planets from brown dwarfs under
consideration is whether the object forms ``as planets do,'' presumably
in a circumstellar disk, or instead as an isolated object. In this
regard, objects like HR~7672B and GL~86B are interesting in that they
resides in the zone of giant planet formation, \ie, $\sim$~5--30~AU
where we know giant planets exist in our own solar system and
theoretical models can most easily form them.  
Giant planets are thought to form at these radii via gas accretion onto
a rocky core.  Subsequent dynamical processes may cause them to migrate
to smaller radii \citep{1996Natur.380..606L, 1997ApJ...482L.211W,
1998ApJ...500..428T}, since the $\lesssim$4~AU planets found by radial
velocity surveys are not thought to have formed in situ.

However, brown dwarfs are believed to be too massive to form by the same
core accretion scenario as giant planets. Instead, they may have
originated, \eg, by fragmentation during initial cloud collapse
\citep{1998bdep.conf..115B}, by instabilities in a very massive disk
\citep{1998ApJ...503..923B, 2000ApJ...540.1091B, 2000ApJ...540L..95P},
or by collisions between protoplanetary disks
\citep{1998Sci...281.2025L}.
Furthermore, brown dwarf companions are not found at small separations,
but we now have strong evidence that they exist at $\sim$15~AU. This
suggest they are immune to the migration process(es) which affect
massive planets.
The discovery of brown dwarfs in the zone of giant planet formation
implies that a diversity of physical processes act to populate the outer
regions of exoplanetary systems.




\acknowledgments

It is a pleasure to thank Bruce Macintosh and Randy Campbell for
beneficial discussions regarding Keck AO data. We thank Mike Brown and
Antonin Bouchez for acquiring the third epoch imaging of HR~7672.  We
are grateful for support from the staffs of Gemini and Keck
Observatories and the University of Hawai'i AO group which made these
observations possible, including Simon Chan, Mark Chun, Pierre Baudoz,
Olivier Guyon, Randy Campbell, David LeMignant, Barbara Schaefer, and
Bob Goodrich.  Gemini Observatory is operated by AURA, Inc., under a
cooperative agreement with the NSF on behalf of the Gemini partnership.
Keck Observatory is operated as a partnership among the California
Institute of Technology, the University of California and the National
Aeronautics and Space Administration, and was made possible by the
generous financial support of the W.M. Keck Foundation.  This paper is
based in part on observations obtained with Hokupa'a/QUIRC, developed
and operated by the University of Hawaii AO Group, with support from the
National Science Foundation.  We thank Sungsoo Kim, Lisa Prato, and Ian
McLean for making the REDSPEC reduction package and documentation
available.  We also thank Aaron Barth, Wayne Landsman, and Ray Sterner
for useful IDL code.  This research has made use of the SIMBAD database
and the VizieR Service \citep{2000A&AS..143...23O} at Centre de
Donn\'ees astronomiques de Strasbourg, and NASA's Astrophysics Data
System Abstract Service.  This work has been supported by NASA grant
NAG5-8299, NSF grants AST95-20443 and AST-9988087, and SUN Microsystems.
M. Liu is grateful for research support from the Beatrice Watson Parrent
Fellowship at the University of Hawai`i.  The authors wish to extend
special thanks to those of Hawaiian ancestry on whose sacred mountain of
Mauna Kea we are privileged to be guests.  Without their generous
hospitality, none of the observations presented herein would have been
possible.

\clearpage





\clearpage
\begin{deluxetable}{llcccccc}
\tablecaption{HR 7672B Astrometry \label{table-astrometry}}
\tablewidth{0pt}
\tabletypesize{\small}

\tablehead{
\colhead{Date (UT)}   & 
\colhead{Telescope} &
\colhead{sep (mas)} &
\colhead{PA (\degs)} &
\multicolumn{2}{c}{observed}  &
\multicolumn{2}{c}{if background object} \\
\colhead{}   & 
\colhead{}   & 
\colhead{}      & 
\colhead{}      & 
\colhead{$\Delta$sep (mas)} &
\colhead{$\Delta$PA (\degs)} &
\colhead{$\Delta$sep (mas)} &
\colhead{$\Delta$PA (\degs)}
}

\startdata
2001 Jun 24 &  Gemini-N &  795 $\pm$ 5 & 157.0 $\pm$ 0.2   &   \nodata   & \nodata      &   \nodata    &   \nodata     \\
2001 Aug 22 &  Keck II  &  786 $\pm$ 6 & 157.9 $\pm$ 0.5   &  $-$9 $\pm$ 8 & 0.9 $\pm$ 0.6  &  $-$31 $\pm$ 5 & \phm{1}$-$6.3 $\pm$ 0.4 \\
2001 Dec 10 &  Keck II  &  794 $\pm$ 5 & 157.3 $\pm$ 0.6   &  $-$1 $\pm$ 7 & 0.3 $\pm$ 0.7  &  $-$61 $\pm$ 5 & $-$19.2 $\pm$ 0.3 \\
\enddata


\end{deluxetable}

\end{document}